\documentclass[twocolumn,prb]{revtex4}
\usepackage{amsfonts}
\usepackage[T1]{fontenc}
\usepackage{amsmath,amsbsy,amssymb,graphicx}
\usepackage{times}

\let\mathbf=\boldsymbol
\def\subsection#1{\paragraph*{\textbf{#1}}}

\begin{document}

\title{Electrically Tunable Quasi-Flat Bands, Conductance and Field Effect
Transistor in Phosphorene}
\author{Motohiko Ezawa}

\affiliation{Department of Applied Physics, University of Tokyo, Hongo 7-3-1, 113-8656, Japan}

\begin{abstract}
Phosphorene, a honeycomb structure of black phosphorus, was isolated
recently. We investigate electric properties of phosphorene nanoribbons based
on the tight-binding model. A prominent feature is the presence of
quasi-flat edge bands entirely detached from the bulk band. We explore the
mechanism of the emergence of the quasi-flat bands analytically and
numerically from the flat bands well known in graphene by a continuous
deformation of a honeycomb lattice. The quasi-flat bands can be controlled
by applying in-plane electric field perpendicular to the ribbon direction.
The conductance is switched off above a critical electric field, which acts
as a field-effect transistor. The critical electric field is
anti-proportional to the width of a nanoribbon. This results will pave a way
toward nanoelectronics based on phosphorene.
\end{abstract}

\maketitle

Graphene is one of the most fascinating material found in this decade\cite%
{NetoRev,KatsText}. The low-energy theory is described by massless Dirac fermions,
which leads to various remarkable electrical properties. In practical
applications to current semiconductor technology, however, we need a finite
band gap in which electrons cannot exist freely. For instance, armchair
graphene nanoribbons have a finite gap depending on the width\cite%
{Fujita,EzawaRibbon}, while bilayer graphene under perpendicular electric
field also has a gap\cite{Bilayer}. It is desirable to find an atomic
monolayer bulk sample which has a finite gap. Silicene is one of the
promising candidates of post graphene materials, which is predicted to be a
quantum spin-Hall insulator\cite{LiuPRL}. A striking property of silicene is
that a topological phase transition is induced by applying electric field\cite%
{EzawaNJP}. Nevertheless, silicene has so far been synthesized only on
metallic surfaces\cite{GLay,Takamura,Kawai}. Another promising candidate is
a transition metal dichalcogenides such as molybdenite\cite{Mak,Zeng,Cao}.

\begin{figure}[t]
\centerline{\includegraphics[width=0.5\textwidth]{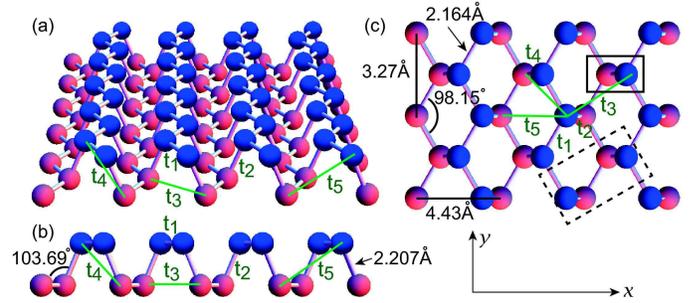}}
\caption{Illustration of the structure and the transfer energy $t_{i}$ of
phosphorene. (a) Bird eye's view. (b) Side view. (c) Top view. The left edge
is zigzag, while the right edge is beard. Red (blue) balls represent
phosphorus atoms in the upper (lower) layer. A dotted (solid) rectangular
denote the unit cell of the 4-band (2-band) model. The parameters of the
unit cell length and angles of bonds are taken from Reference\protect\cite%
{Gomez}.}
\label{FigIllust}
\end{figure}

A new comer challenges the race of the post-graphene materials. That is
phosphorene, a honeycomb structure of phosphorus. It has been successfully
obtained in the laboratory\cite{Li,Liu,Xia,Gomez,Koenig} and revealed a
great potential in applications to electronics. Black phosphorus is a
layered material where individual atomic layers are stacked together by Van
der Waals interactions. Just as graphene can be isolated by peeling
graphite, phosphorene can be similarly isolated from black phosphorus by the
mechanical exfoliation method. As a key structure it is not planer but
puckered due to the $sp^{3}$\ hybridization, as shown in Fig.\ref{FigIllust}%
. There are already several works based on first-principle calculations\cite%
{Peng,Tran,Qiao,Fei}. The tight-binding model was proposed\cite{Kats} only
recently by including the transfer energy $t_{i}$ over the 5 neighbor
hopping sites ($i=1,2,\cdots ,5$), as illustrated in Fig.\ref{FigIllust}.

The aim of this paper is to investigate the physics of phosphorene
nanoribbons based on the tight-binding model. The tight-binding model is
essential to make a deeper understanding of the system, which is not
attained by first-principle calculations and low-energy effective theory. Aa
a striking property of phosphorene nanoribbon, we demonstrate the
presence of a quasi-flat edge band which is entirely detached from the bulk
band. We explore the band structure of phosphorene nanoribbon numerically
and analytically from that of graphene nanoribbon as a continuous
deformation of the honeycomb lattice by changing the transfer energy
parameters $t_{i}$. The graphene is well explained in terms of electron
hopping between the first neighbor sites with $t_{1}=t_{2}\neq 0$ and $%
t_{3}=t_{4}=t_{5}=0$, where the presence of flat bands is well known\cite%
{Fujita,EzawaRibbon}. We follow the fate of the flat bands by changing these
parameters. The essential roles is played by the ratio $%
t_{2}/t_{1}$: The flat bands are detached from the bulk band when the ratio
is $2$, while the flat bands are bent by the term $t_{4}$.

It is an exciting problem to control the band structure externally. We may
change the band gap by applying external electric field $E_{z}$
perpendicular to the phosphorene sheet with the use of the puckered
structure. Although we can control the band gap, the amount of the
controlled gap is very tiny compared to the large band gap, that
is, $\Delta =1.52$eV since the height of the puckered structure $\ell $ is
the order of nm. On the other hand, the quasi-flat edge band is found to be
sensible to external electric field $E_{x}$ parallel to the phosphorene
sheet. This is because a large potential difference ($\varpropto WE_{x}$) is
possible between the two edges if the width $W$ of the nanoribbon is large
enough. We propose a field-effect transistor driven by in-plane electric
field with the use of the quasi-flat edge bands. The conductance is shown to
be either $0$ or $2e^{2}/h$ with respect to the in-plane electric field. If
the nanoribbon width is $1\mu m$, the critical electric field is given by $%
E_{\text{cr}}=0.15$meV/nm. This is experimentally feasible.

\begin{figure}[t]
\centerline{\includegraphics[width=0.48\textwidth]{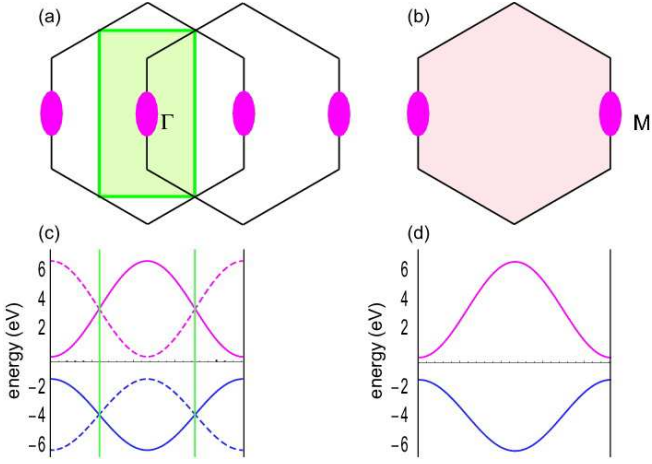}}
\caption{Brillouin zones and energy spectra of the 4-band and 2-band models
of phosphorene. (a) The Brillouin zone is a rectangular in the 4-band model,
which is constructed from two copies of the hexagonal Brillouin zone of the
2-band model. A magenta oval denotes a Dirac cone present at the $\Gamma $
point. (b) The Brillouin zone is a hexagonal in the 2-band model. A magenta
oval denotes a Dirac cone present at the $M$ point. (c) The band structure
of the 4-band model, which is constructed from two copies of that of the
2-band model. (d) The band structure of the 2-band model. }
\label{FigBrill}
\end{figure}

\subsection{4-band tight-binding model.}

The unit cell of phosphorene contains 4 phosphorus atoms, where two
phosphorus exist in the upper layer and the other two phosphorus exist in
the lower layer. The tight-binding model of phosphorene was recently proposed%
\cite{Kats} and is given by
\begin{equation}
H_{4}=\sum_{\left\langle i,j\right\rangle }t_{ij}c_{i}^{\dagger }c_{j},
\end{equation}%
where summation runs over the lattice sites, $t_{ij}$ is the transfer energy
between $i$th and $j$th sites, and $c_{i}^{\dagger }$ ($c_{j}$) is the
creation (annihilation) operator of electrons at site $i$ ($j$). It has been
shown that it is enough to take $5$ hopping links, as illustrated in Fig.\ref%
{FigIllust}. The transfer energy explicitly reads as $t_{1}=-1.220$eV, $%
t_{2}=3.665$eV, $t_{3}=-0.205$eV, $t_{4}=-0.105$eV, $t_{5}=-0.055$eV for
these links.

In the momentum representation the 4-band Hamiltonian reads as $H_{4}=\sum_{%
\mathbf{k}}c^{\dagger }(\mathbf{k})\hat{H}_{4}(\mathbf{k})c(\mathbf{k})$ with%
\begin{equation}
\hat{H}_{4}=\left(
\begin{array}{cccc}
0 & f_{1}+f_{3} & f_{4} & f_{2}+f_{5} \\
f_{1}^{\ast }+f_{3}^{\ast } & 0 & f_{2} & f_{4} \\
f_{4}^{\ast } & f_{2}^{\ast } & 0 & f_{1}+f_{3} \\
f_{2}^{\ast }+f_{5}^{\ast } & f_{4}^{\ast } & f_{1}^{\ast }+f_{3}^{\ast } & 0%
\end{array}%
\right) ,  \label{Hamil4}
\end{equation}%
where
\begin{align}
f_{1}=& 2t_{1}e^{ik_{x}/2}\cos \frac{\sqrt{3}}{2}k_{y},\quad
f_{2}=t_{2}e^{-ik_{x}},  \notag \\
f_{3}=& 2t_{3}e^{-5ik_{x}/2}\cos \frac{\sqrt{3}}{2}k_{y},  \notag \\
f_{4}=& 4t_{4}\cos \frac{3}{2}k_{x}\cos \frac{\sqrt{3}}{2}k_{y},\quad
f_{5}=t_{5}e^{2ik_{x}}.  \label{f-value}
\end{align}%
We show the Brillouin zone and the energy spectrum of the tight-binding
model in Fig.\ref{FigBrill}(a) and (c), respectively.

\subsection{2-band tight-binding model.}

We are able to reduce the 4-band model to the 2-band model due to the $C_{2h}
$ point group invariance. We focus on a blue point (atom in upper layer) and
view other lattice points in the crystal structure (Fig.\ref{FigIllust}). We
also focus on a red point (atom in lower layer) and view other lattice
points. As far as the transfer energy is concerned, the two views are
identical. Namely, we may ignore the color of each point. Hence, instead of
the unit cell containing 4 points, it is enough to consider the unit cell
containing only 2 points. This reduction makes our analytical study
considerably simple.

The 2-band model is given by $H_{2}=\sum_{\mathbf{k}}c^{\dagger }(\mathbf{k})%
\hat{H}_{2}(\mathbf{k})c(\mathbf{k})$ with
\begin{equation}
\hat{H}_{2}=\left(
\begin{array}{cc}
f_{4} & f_{1}+f_{2}+f_{3}+f_{5} \\
f_{1}^{\ast }+f_{2}^{\ast }+f_{3}^{\ast }+f_{5}^{\ast } & f_{4}%
\end{array}%
\right) .  \label{Hamil2}
\end{equation}%
The rectangular Brillouin zone of the 4-band model is constructed by folding
the hexagonal Brillouin zone of the 2-band model, as illustrated in Fig.\ref%
{FigBrill}(a).

The equivalence between the two models (\ref{Hamil4}) and (\ref{Hamil2}) is
verified as follows. We have explicitly shown the energy spectra of the
4-band model (\ref{Hamil4}) and the 2-band model (\ref{Hamil2}) in Fig.\ref%
{FigBrill}(c) and (d), respectively. It is demonstrated that the energy
spectrum of the 4-band model is constructed from that of the 2-band model:
The two bands are precisely common between the two models, while the extra
two bands in the 4-band model are obtained simply by shifting the two bands
of the 2-band model, as dictated by the folding of the Brillouin zone.

\begin{figure}[t]
\centerline{\includegraphics[width=0.5\textwidth]{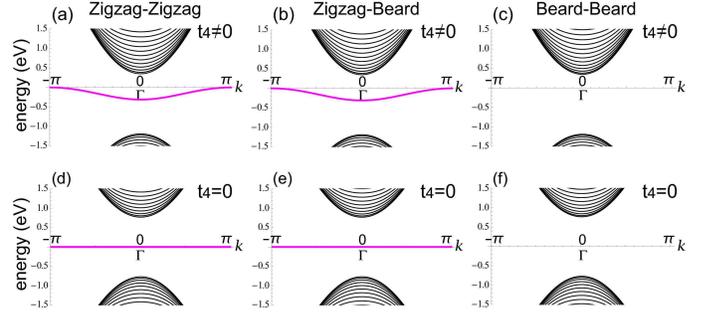}}
\caption{Band structure of phosphorene nanoribbons when the transfer energy $%
t_4$ is nonzero and zero. (a,d) Both edges are zigzag. (b,e) One edge is
zigzag and the other edge is beard. (c,f) Both edges are beard. The
quasi-flat edge mode emerges for (a) and (b). When we set $t_{4}=0$, the
quasi-flat band becomes perfectly flat band. Flat and quasi-flat Edge states
are marked in magenta.}
\label{FigRibbon}
\end{figure}

\subsection{Phosphorene nanoribbons.}

We investigate the band structure of a phosphorene nanoribbon placed along
the $y$ direction (Fig.\ref{FigIllust}). We study zigzag and beard edges.
There are three types of nanoribbons, whose edges are (a) both zigzag, (b)
zigzag and beard, (c) both beard. We show their band structures in Fig.\ref%
{FigRibbon}(a), (b) and (c), respectively.

A prominent feature is the presence of the edge modes isolated from the bulk
modes found in Fig.\ref{FigRibbon}(a) and (b). They comprise a quasi-flat
band. It is doubly degenerate for a zigzag-zigzag nanoribbon, and
nondegenerate for a zigzag-beard nanoribbon, while it is absent in a beard-beard nanoribbon [Fig.\ref{FigRibbon}(c)].

Let us explore the mechanism how such an isolated quasi-flat band appears in phosphorene.
By diagonalizing the Hamiltonian, the energy spectrum reads
\begin{equation}
E(\mathbf{k})=f_{4}\pm \left\vert f_{1}+f_{2}+f_{3}+f_{5}\right\vert .
\label{EnergSpect2}
\end{equation}%
The band gap is given by%
\begin{equation}
\Delta =4t_{1}+2t_{2}+4t_{3}+2t_{5}=1.52\text{eV}.
\end{equation}%
The asymmetry between the positive and negative energies arises from the $%
f_{4}$ term. Then it is interesting to see what would happen when we set $%
t_{4}=0$. We show the band structures in Fig.\ref{FigRibbon}(d), (e) and (f)
for the three types of nanoribbons, where the quasi-flat edge modes are
found to become perfectly flat. Consequently it is enough to show the
emergence of the flat band by studying the model with $t_{4}=0$. Furthermore
it is a good approximation to set $t_{3}=t_{5}=0$, since the transfer
energies $t_{1}$ and $t_{2}$ are much larger than the others. Indeed we have
checked numerically that no qualitative difference is induced by this
approximation.

\subsection{Flat bands in anisotropic honeycomb lattice.}

To explore the origin of the isolated quasi-flat band we analyze the
anisotropic honeycomb-lattice model, which is described by the Hamiltonian (%
\ref{Hamil2}) by setting $f_{3}=f_{4}=f_{5}=0$. This Hamiltonian is well
studied in the context of graphene and optical lattice\cite%
{Pereira,Monta,Wunsch,Tar}. The energy spectrum reads%
\begin{equation}
E=\sqrt{t_{2}^{2}+4\left( t_{1}^{2}+t_{1}t_{2}\cos \frac{3}{2}k_{x}\right)
\cos \frac{\sqrt{3}}{2}k_{y}},  \label{EnergSpec}
\end{equation}%
which implies the existence of two Dirac cones at
\begin{equation}
k_{x}=\pm \arctan (\sqrt{4t_{1}^{2}-t_{2}^{2}}/t_{2}),\quad k_{y}=0,
\label{DiracPoint}
\end{equation}%
for $\left\vert t_{2}\right\vert <2\left\vert t_{1}\right\vert $.

\begin{figure}[t]
\centerline{\includegraphics[width=0.5\textwidth]{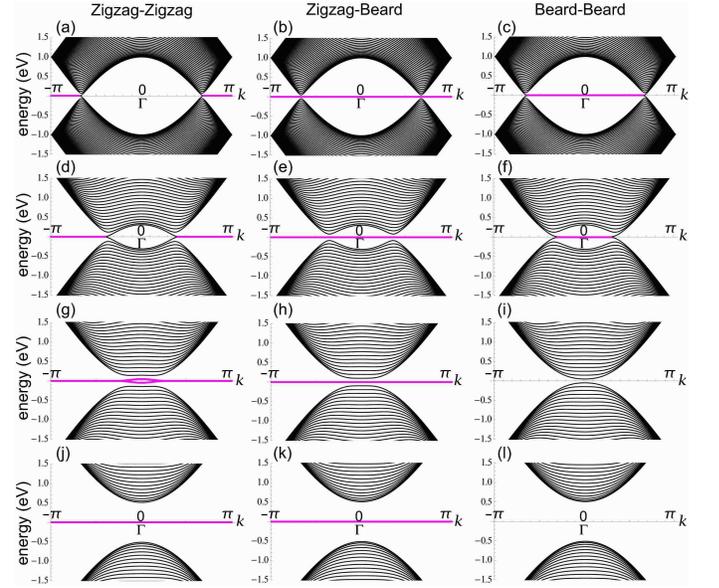}}
\caption{Band structure anisotropic honeycomb nanoribbons. The flat edge
states are marked in magenta. We have set $t_{1}=-1$ and $%
t_{3}=t_{4}=t_{5}=0 $. We have also set (a,b,c) $t_{2}=1$, (d,e,f) $%
t_{2}=1.7 $, (g,h,i) $t_{2}=2$, (j,k,l) $t_{2}=2.5$. The unit cell contains $%
144$ atoms.}
\label{FigMerge}
\end{figure}

We now study the change of the band structure of nanoribbon
by a continuous deformation of the honeycomb lattice,
starting that of graphene.
We show the band structure with (a) the zigzag-zigzag edges, (b) the zigzag-beard edges,
and (c) the beard-beard edges in Fig.\ref{FigMerge} for typical values of $%
t_{1}$ and $t_{2}$.

(i) We start with the isotropic case $t_{1}=t_{2}$, where the energy
spectrum (\ref{EnergSpec}) becomes that of graphene with two Dirac cones at
the $K$ and $K^{\prime }$ points. The perfect flat band connects the $K$ and
$K^{\prime }$ points, that is, it lies for (a) $-\pi \leq k\leq -\frac{2}{3}%
\pi $ and $\frac{2}{3}\pi \leq k\leq \pi $; (b) $-\pi \leq k\leq \pi $; (c) $%
-\frac{2}{3}\pi \leq k\leq \frac{2}{3}\pi $. It is attached to the bulk
band. See Fig.\ref{FigMerge}(a),(b),(c). The topological origin of flat
bands in graphene has been thoroughly discussed\cite{Hatsugai}.

(ii) As we increase $t_{2}$ but keeping $t_{1}$ fixed, the two Dirac points
move towards the $\Gamma $ point ($k=0$), as is clear from (\ref{DiracPoint}%
). The flat band keeps to be present between the two Dirac points. See Fig.%
\ref{FigMerge}(d),(e),(f).

(iii) At $\left\vert t_{2}\right\vert =2\left\vert t_{1}\right\vert $, the
two Dirac points merge into one Dirac point at the $\Gamma $ point, as
implied by (\ref{DiracPoint}). The flat band touches the bulk band at the $%
\Gamma $ point for the zigzag-zigzag nanoribbon and the zigzag-beard
nanoribbon, but disappears from the beard-beard nanoribbon. See Fig.\ref%
{FigMerge}(g),(h),(i).

(iv) For $\left\vert t_{2}\right\vert >2\left\vert t_{1}\right\vert $, the
bulk band shifts away from the Fermi level, as follows from (\ref{EnergSpec}%
). The flat band is disconnected from the bulk band for the zigzag-zigzag
nanoribbon and the zigzag-beard nanoribbon, where it extends over all region
$-\pi \leq k\leq \pi $. On the other hand, the edge band becomes a part of
the bulk band and disappears from the Fermi level for the beard-beard
nanoribbon. See Fig.\ref{FigMerge}(j),(k),(l).

\subsection{Energy spectrum of quasi-flat bands.}

We have explained how the flat band appears in the anisotropic
honeycomb-lattice model. The flat band is bent into the quasi-flat band by
switching on the transfer interaction $t_{4}$.

We can derive the energy spectrum of the quasi-flat band perturbatively.
We construct an analytic form of the wave function at the zero-energy state
in the anisotropic honeycomb-lattice model ($t_{3}=t_{4}=t_{5}=0$). By
solving the Hamiltonian matrix recursively from the outer most cite, we
obtain the analytic form of the local density of states of the wave function
for odd cite $j$,
\begin{equation}
|\psi (j)|=\alpha ^{j}\sqrt{1-\alpha ^{2}},  \label{wave}
\end{equation}%
with $\alpha =2|t_{1}|(\cos \frac{k}{2})/|t_{2}|$. The wave function is zero
for even cite. It is perfectly localized at the outer most cite when $k=\pi $%
, and describes the flat band. With the use of this wave function, the
energy spectrum $E_{\text{qf}}(k)$ of the quasi-flat band is\ estimated
perturbatively by taking the expectation value of the $t_{4}$ term as
\begin{equation}
E_{\text{qf}}(k)=-\frac{4t_{1}t_{4}}{t_{2}}(1+\cos k)\quad \text{eV},
\label{DispQF}
\end{equation}%
where $4t_{1}t_{4}/t_{2}=0.14$. On the other hand we numerically obtain $%
E(0)=-0.301$eV. The agreement is excellent.

\subsection{Phosphorene nanoribbons with in-plane electric field.}

\begin{figure}[t]
\centerline{\includegraphics[width=0.4\textwidth]{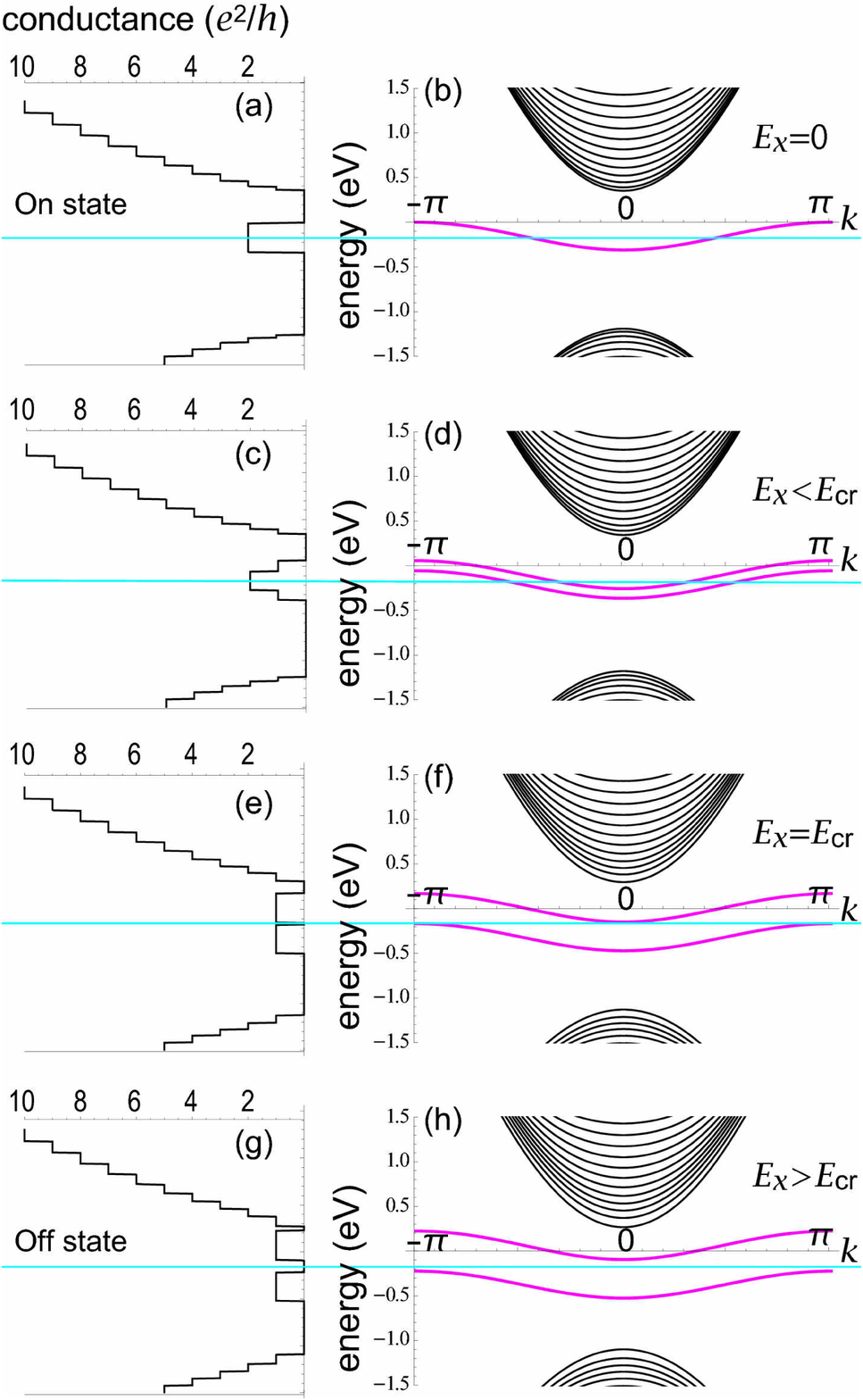}}
\caption{Band structure and conductance in unit of $e^{2}/h$ of phosphorene
nanoribbons with the both edges being zigzag in the presence of in-plane electric field $E_x$.
(a,b) $E_{x}=0$, (c,d) $E_{x}=0.5$meV/nm, (e,f) $E_{x}=1.5$%
meV/nm, (g,h) $E_{x}=2.0$meV/nm. Magenta curves represent the quasi-flat
bands, while cyan lines represent the Fermi energy. The unit cell contains $%
144$ atoms.}
\label{FigIE}
\end{figure}

It is an intriguing problem to control the band structure externally. We may
try to change the band gap by applying external electric field $E_{z}$
perpendicular to the phosphorene sheet. We find the band gap to behave as%
\begin{equation}
\Delta =1.52+0.28\left( \ell E_{z}\right) ^{2}\text{eV},  \label{GapEz}
\end{equation}%
where $\ell $ is the separation between the upper and lower layers. The gap
becomes simply larger when we apply $E_{z}$. Furthermore, in practical
applications, it needs very large electric field since $\ell $\ is the order
of nm.

On the other hand, it is possible to make a significant change of the
quasi-flat edge band by applying external electric field parallel to the
phosphorene sheet. Let us apply electric field $E_{x}$ into the $x$
direction of a nanoribbon with zigzag-zigzag edges. A large potential
difference ($\varpropto WE_{x}$) is possible between the two edges if the
width $W$ of the nanoribbon is large enough. This potential difference
resolves the degeneracy of the two edge modes, shifting one edge mode
upwardly and the other downwardly without changing their shapes. We present
the resultant band structures in Fig.\ref{FigIE} for typical values of $%
E_{x} $.

\subsection{Field-effect transistor.}

Electric current may flow along the edge. We show the conductance in Fig.\ref%
{FigIE}. Without in-plane electric field, the conductance at the Fermi
energy is $2e^{2}/h$ since there is a two-fold degenerate quasi-flat band.
Above the critical electric field, the conductance becomes $0$ since the
quasi-flat band splits perfectly. This acts as a field-effect transistor
driven by in-plane electric field. The critical electric field is
anti-proportional to the width $W$.

We derive the critical electric field $E_{\text{cr}}$. The energy shift at $%
k=\pi $ is given by $\Delta E\left( \pi \right) =\pm WE_{x}$ since the wave
function is perfectly localized at the outer most edge cite. In general the
energy shift is given by
\begin{equation}
\Delta E\left( k\right) =\pm \left( W-\frac{\alpha ^{2}}{1-\alpha ^{2}}%
\right) E_{x}.  \label{EnergShift}
\end{equation}%
It is well approximated by $\Delta E\left( k\right) =\pm WE_{x}$ for wide
nanoribbons. The conductance is written as
\begin{align}
& \frac{e^{2}}{h}[\theta (E-\varepsilon _{0}-\left\vert \Delta E\left(
0\right) \right\vert )-\theta (E-\left\vert \Delta E\left( \pi \right)
\right\vert )  \notag \\
& \quad \quad +\theta (E-\varepsilon _{0}+\left\vert \Delta E\left( 0\right)
\right\vert )-\theta (E+\left\vert \Delta E\left( \pi \right) \right\vert )],
\end{align}%
where $\theta (x)$ is the step function $\theta =1$ for $x>0$ and $\theta =0$
for $x<0$. The critical electric field is determined as
\begin{equation}
E_{\text{cr}}=|\varepsilon _{0}|/(2W)
\end{equation}%
If the nanoribbon width is $1\mu m$, the critical electric field is given by
$E_{\text{cr}}=0.15$meV/nm. This is experimentally feasible.

\subsection{Discussion.}
We have analyzed the band structure of phosphorene nanoribbons based on the
tight-binding model, and demonstrated the presence of quasi-flat edge modes
entirely detached from the bulk band. Starting from the well-known structure
of graphene, we have explained the mechanism how such edge modes emerge by a
continuous deformation of the honeycomb lattice. The conductance due to the
quasi-flat edge modes is quantized to be either $0$ or $2e^{2}/h$ with
respect to the in-plane electric field $E_{x}$. The critical electric field
is $E_{\text{cr}}=0.15$meV/nm for a nanoribbon with width $1\mu m$. A
field-effect transistor is possible with the use of this property. We may
also expect a similar structure made of arsenic and antimony, which should
be called as "arsenene" and "antimonene". The electronic properties will be
explained simply by setting the transfer energies appropriately.

\subsection{Acknowledgements}
The author thanks the support by the Grants-in-Aid for Scientific
Research from the Ministry of Education, Science, Sports and Culture No.
25400317. M. E. is very much grateful to N. Nagaosa for many helpful
discussions on the subject.

\newpage
\section*{Supporting Materials}

\subsection{Wave function of edge state.}

\quad We construct an analytic form of the wave function at the zero-energy
state in the anisotropic honeycomb-lattice model ($t_{3}=t_{4}=t_{5}=0$) as
follows. We label the wave function of the atom on the outer most cite as $%
\psi _{1}$, and that of the atom next to it as $\psi _{2}$, and as so on. The
total wave function is $\psi =\left\{ \psi _{1},\psi _{2},\cdots ,\psi
_{N}\right\} $ if there are $N$ atoms across the nanoribbon. The Hamiltonian
is explicitly written as

\begin{equation}
H=\left(
\begin{array}{ccccc}
0 & t_{1}g^{\ast } & 0 & 0 & \cdots \\
t_{1}g & 0 & t_{2} & 0 & \cdots \\
0 & t_{2} & 0 & t_{1}g^{\ast } & \cdots \\
0 & 0 & t_{1}g & 0 & \cdots \\
\cdots & \cdots & \cdots & \cdots & \cdots%
\end{array}%
\right) ,
\end{equation}%
with $g=1+e^{ik}$. The eigenvalue problem $H\psi =0$ is trivially solved,
yielding $\psi _{2n}=0$ and $\psi _{2n+1}=[t_{1}\left( 1+e^{ik}\right)
/t_{2}]^{n}\psi _{1}$.

The dispersion of the quasi-flat band is determined as
\begin{align}
E_{\text{qf}}(k) =&\sum_{n=0}^{\infty }t_{4}(1+e^{-ik})\psi _{n}^{\ast }\psi
_{n+2}+t_{4}(1+e^{ik})\left( \psi _{n+2}^{\ast }\psi _{n}\right)  \notag \\
=&-\frac{4t_{1}t_{4}}{t_{2}}(1+\cos k),
\end{align}%
which is (\ref{DispQF}) in the text. The energy shift due to the in-plane
electric field is given by%
\begin{equation}
\sum_{n=0}^{\infty }\left( W-n\right) \left\vert \psi _{n}\right\vert
^{2}=\left( W-\frac{\alpha ^{2}}{1-\alpha ^{2}}\right) E_{x},
\end{equation}%
which is (\ref{EnergShift}) in the text.

\subsection{Electric field.}

\quad We apply electric field $E_{z}$ perpendicular to the sheet. It is
necessary to use the 4-band tight-binding model, since the electric field
breaks the $C_{2h}$ point group invariance. Namely, the upper and lower
layers are distinguished by the electric field. We introduce $E_{z}$
into the Hamiltonian (\ref{Hamil4}),
\begin{equation}
\hat{H}_{4}(E_{z})=\hat{H}_{4}+\text{diag.}(\ell E_{z},\ell E_{z},-\ell
E_{z},-\ell E_{z}).
\end{equation}%
By diagonalizing this Hamiltonian at $\mathbf{k}=0$, the band gap
is found to be
\begin{equation}
\Delta =\sum_{s=\pm 1}\sqrt{\left( t_{2}+s4t_{4}+t_{5}\right) ^{2}+\left(
\ell E_{z}\right) ^{2}}+4\left( t_{1}+t_{3}\right) ,
\end{equation}%
which yields numerically (\ref{GapEz}) in the text.

\subsection{Low-energy theory.}

\quad In the vicinity of the $\Gamma $ point, we make a Tayler expansion and
obtain%
\begin{align}
f_{1}=& t_{1}\left( 2+ik_{x}-\frac{1}{4}k_{x}^{2}-\frac{3}{4}%
k_{y}^{2}\right) ,  \notag \\
f_{2}=& t_{2}\left( 1-ik_{x}-\frac{1}{2}k_{x}^{2}\right) ,  \notag \\
f_{3}=& t_{3}\left( 2-5ik_{x}-\frac{25}{4}k_{x}^{2}-\frac{3}{4}%
k_{y}^{2}\right) ,  \notag \\
f_{4}=& t_{4}\left( 4-\frac{9}{2}k_{x}^{2}-\frac{3}{2}k_{y}^{2}\right) ,
\notag \\
f_{5}=& t_{5}\left( 1+2ik_{x}-2k_{x}^{2}\right) .
\end{align}%
The Hamiltonian (\ref{Hamil2}) reads%
\begin{equation}
\hat{H}_{2}=f_{4}+\left( \varepsilon +\alpha k_{x}^{2}+\beta
k_{y}^{2}\right) \tau _{x}+\gamma k_{x}\tau _{y},
\end{equation}%
with the Pauli matrices $\mathbf{\tau }$, where%
\begin{align}
\varepsilon =& 2t_{1}+t_{2}+2t_{3}+t_{5}=0.76\text{eV,}  \notag \\
\alpha =& -\frac{1}{4}t_{1}-\frac{1}{2}t_{2}-\frac{25}{4}t_{3}-\frac{9}{2}%
t_{4}-2t_{5}=0.336\text{eV,}  \notag \\
\beta =& -\frac{3}{4}t_{1}+\frac{3}{4}t_{3}+\frac{3}{2}t_{4}=0.604\text{eV,}
\notag \\
\gamma =& -t_{1}+t_{2}+5t_{3}-2t_{5}=3.97\text{eV.}
\end{align}%
Hence the dispersion is linear in the $k_{y}$ direction but parabolic in the
$k_{y}$ direction. The low-energy Hamiltonian agrees with the previous result%
\cite{Rodin} with a rotation of the Pauli matrices $\tau _{x}\mapsto \tau
_{z}$ and $\tau _{y}\mapsto \tau _{x}$.

\subsection{Conductance.}

\quad In terms of single-particle Green's functions, the low-bias
conductance $\sigma (E)$ at the Fermi energy $E$ is given by\cite{Datta}
\begin{equation}
\sigma (E)=(e^{2}/h)\text{Tr}[\Gamma _{\text{L}}(E)G_{\text{D}}^{\dag
}(E)\Gamma _{\text{R}}(E)G_{\text{D}}(E)],
\end{equation}%
where $\Gamma _{\text{R(L)}}(E)=i[\Sigma _{\text{R(L)}}(E)-\Sigma _{\text{%
R(L)}}^{\dag }(E)]$ with the self-energies $\Sigma _{\text{L}}(E)$ and $%
\Sigma _{\text{R}}(E)$, and%
\begin{equation}
G_{\text{D}}(E)=[E-H_{\text{D}}-\Sigma _{\text{L}}(E)-\Sigma _{\text{R}%
}(E)]^{-1},  \label{StepA}
\end{equation}%
with the Hamiltonian $H_{\text{D}}$ for the device region. The self-energy $%
\Sigma _{\text{L(R)}}(E)$ describes the effect of the electrode on the
electronic structure of the device, whose the real part results in a shift
of the device levels whereas the imaginary part provides a life time. It is
to be calculated numerically\cite{Sancho,Rojas,Nikolic,EzawaAPL}. We have
used this formula to derive the conductance in Fig.\ref{FigIE}.

\end{document}